\documentclass[12pt,preprint]{aastex}
\usepackage{graphicx}
\usepackage{subfigure}
\usepackage{lscape}
\usepackage{cite}

\slugcomment{}
\shorttitle{Rotating Gas Surrounding HII Regions}
\shortauthors{Klaassen et al.}

\begin{document}

\title{Rotation of the Warm Molecular Gas Surrounding Ultracompact HII Regions.}

\author{P. D. Klaassen\altaffilmark{1,2}, C. D. Wilson\altaffilmark{1}, E. R. Keto\altaffilmark{3} \& Q. Zhang\altaffilmark{3}}
\altaffiltext{1}{Dept. of Physics and Astronomy, McMaster University, Hamilton, Canada}
\altaffiltext{2}{Currently at the European Southern Observatory (ESO), Garching, Germany (ALMA Fellow)}
\altaffiltext{3}{Harvard-Smithsonian Center for Astrophysics, Cambridge, USA}
\email{pklaasse@eso.org}

\begin{abstract}

We present molecular line and 1.4 mm continuum observations towards five massive star forming regions at arcsecond resolution using the Submillimeter Array (SMA).  We find that the warm molecular gas surrounding each HII region (as traced by SO$_2$ and OCS) appears to be undergoing bulk rotation.  From the molecular line emission and thermal component of the continuum emission, we independently derived gas masses for each region which are consistent with each other.  From the free-free component of the continuum emission we estimate the minimum stellar mass required to power the HII region and find that this mass, when added to the derived gas mass, is a significant fraction of the dynamical mass for that region.  
\end{abstract}

\keywords{Stars: Formation -- Molecular data -- ISM: Kinematics and Dynamics -- sub-millimeter }

\section{Introduction}
\label{sec:intro}

As a core within an overdense region of a molecular cloud begins to collapse, its angular momentum causes the rotating core to speed up, often forming an accretion disk.   Until recently, it was suggested that for high mass stars, accretion must be halted at the onset of an HII region, that the strong outward pressures must reverse an accretion flow \citep[i.e.][]{GL99,Churchwell02}. However, recent models which include the effects of gravity have suggested that accretion can continue \citep{Keto03,Keto07} and observational support has followed for infall and accretion in both ionized gas \citep{KW06} and molecular gas \citep{Sollins05,KW07,KW08} in the presence of HII regions.  If we are to understand high mass star formation, we must first understand the dynamics of their natal environments both before and after the formation of an HII region.

If analogies can be drawn between low mass and high mass star formation, we would expect this continued accretion flow to occur though a disk \citep[i.e.][]{Zhang05,Cesaroni07}.  If these disks exist, we cannot yet observe them except in the closest sources.  However, using submillimeter telescopes, we can observe the warm molecular gas around ultracompact HII (UCHII) regions to determine whether there is large scale rotation which could translate onto smaller scales on which disks may exist.  Continuum studies at high resolution may show flattened structures perpendicular to outflows \citep[i.e][]{Gibb07,Zhang07}; however, without observing the kinematics of the gas, it is unclear whether these structures are indeed rotating. When and how accretion disks form is still fairly uncertain for low mass stars, and their presence in high mass star forming regions is still often debated. However, there is a growing number of high mass star forming regions in which disks have been detected (such as G24A1 \citep{Beltran06}, IRAS 20126 \citep{Cesaroni05}, and IRAS 18089 \citep{Beuther08}) all of which have yet to form HII regions.  In addition to these detections, rotating structures seen in NH$_3$ have also been observed in a few sources at high resolution with the VLA \citep[i.e.][]{Sollins05,Zhang97,Keto88,Keto87,Zhang98}.    Here, we present observations of warm gas tracers (SO$_2$ and OCS) towards five high mass star forming regions which are suggestive of bulk rotation in the gas surrounding each UCHII region.

At 220 GHz, the SMA\footnote{The Submillimeter Array is a joint project between the Smithsonian Astrophysical Observatory and the Academia Sinica Institute of Astronomy and Astrophysics, and is funded by the Smithsonian Institution and the Academia Sinica.} in its extended configuration is only sensitive to structures smaller than $\sim$ 10$''$. Thus, at the average distance to the massive star forming regions studied here (6.4 kpc), we can only observe structures smaller than $\sim$ 64000 AU, and our data are ideally suited to probe the warm molecular gas surrounding HII regions.  

In Section \ref{sec:observations} below, we briefly introduce our observations, which come from the same dataset as those presented in \citet{KZK08}.  In Section \ref{sec:results}, we present our analysis of the velocity gradients in SO$_2$ and OCS and the dynamical masses we derive for each region.  In Section \ref{sec:other_mass} we present our other mass determination methods.  In Section \ref{subsec:temperature} we derive gas temperatures based on the J=12-11 K ladder of CH$_3$CN.  We then use these derived temperatures to estimate the gas mass responsible for the SO$_2$ and OCS line emission (Section \ref{subsec:emission_mass}).  From the thermal component of the continuum emission we derive a gas mass based on the emission from the warm dust (Section \ref{subsec:continuum_mass}).  In Section \ref{subsec:stellar_mass} we present stellar mass estimates based on the free-free continuum emission from these sources.  In Section \ref{sec:discussion}, we discuss the implications of these different mass determinations for each source, and conclude.

\section{Observations}
\label{sec:observations}

The observations presented here were obtained at the SMA in its extended configuration in the fall of 2005. Source names, pointing centers and source offsets from the pointing centers are given in Table \ref{tab:observe}.  H30$\alpha$ was centered in chunk 21 of the upper sideband, and the observations were made with a spectral resolution of 0.53 km s$^{-1}$ over the entire 2 GHz bandwidth in each sideband.  This correlator setup allowed for the simultaneous observation of OCS (J=19-18 at 231.061 GHz) in the upper sideband as well as SO$_2$ (J=11$_{1,11}$-10$_{0,10}$ at 221.965 GHz) and CH$_3$CN (J=12-11) in the lower sideband.  The K components of the CH$_3$CN molecule up to K=8 were observed towards each source.    CO and $^{13}$CO (J=2-1) were also observed; however they are presented in a separate paper (Klaassen et al. in prep).  Data reduction was completed using MIRIAD in both the uv and image planes. Also presented here is the 1.4 mm continuum emission, which was derived from the line free portions of the large bandwidth (2GHz) of the lower sideband of these SMA observations. The distances to each source and rms noises of these observations are given in Table \ref{tab:observe}. The distances used here for the regions in W51 and NGC 7538 IRS 1 are smaller than those used in \citet{Keto08}, and were taken from the parallax measurements of \citet{Xu09} and \citet{Moscadelli09} respectively. Below, an absolute calibration uncertainty of 20\% is assumed, but not propagated through our calculations. When uncertainties are listed, they arise from the rms in the map, propagated through the equations, unless stated otherwise.

\section{Systematic Velocity Gradients in SO$_2$ and OCS}
\label{sec:results}

\begin{figure}
\includegraphics[width=0.75\textwidth,angle=90]{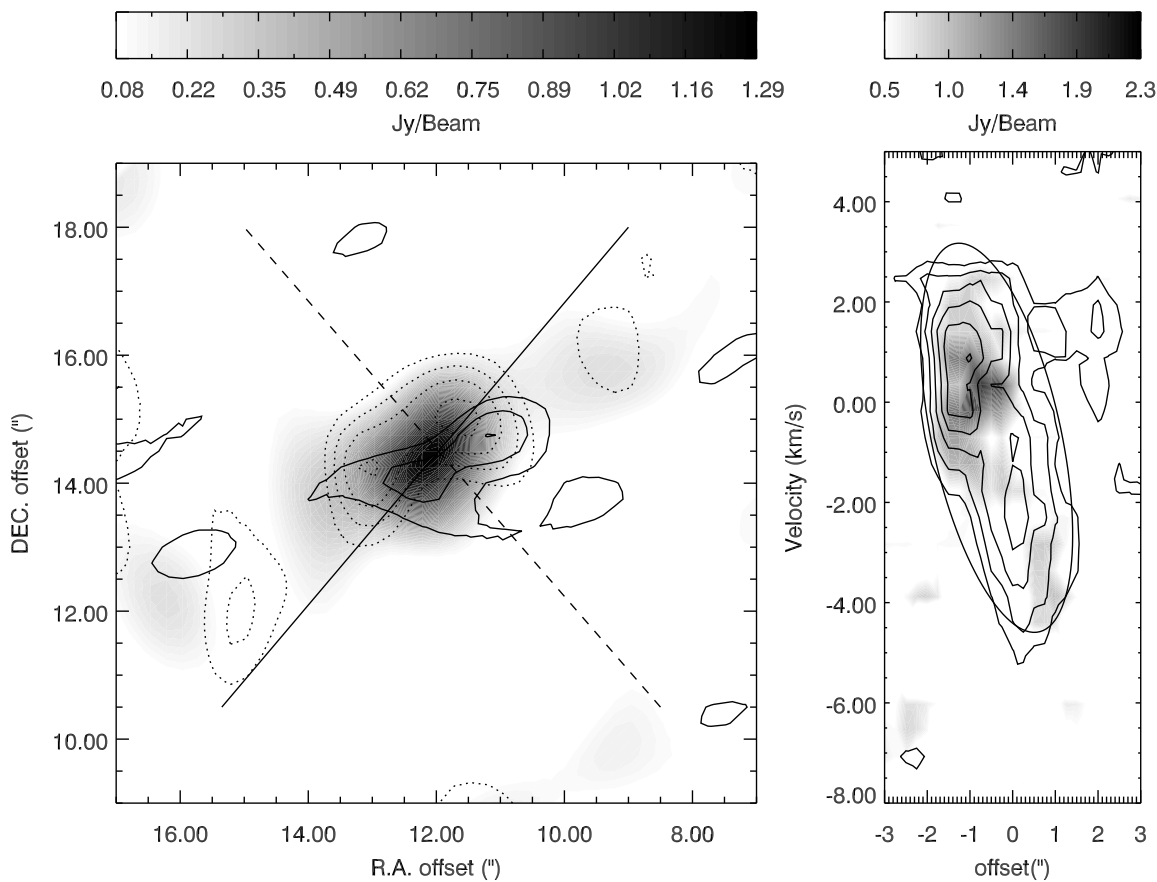}
\caption{{\bf (G10.6-0.4)} {\it Left:} 1.4 mm continuum emission cut at 3$\sigma$ is shown in grayscale with SO$_2$ and OCS emission shown with solid and dotted contours (respectively).  The solid line shows the cut used for the PV diagram shown on the right while the dashed line shows the direction of the outflow as observed in CO (Klaassen et al. in prep).  Contours begin at  3$\sigma$ for each integrated intensity map, and increase in increments of 2$\sigma$.{\it Right:} Position-Velocity (PV) diagram of OCS (greyscale) and SO$_2$ (contours) emission along the cut shown in the left panel. Again, the greyscale and contours both start at 3$\sigma$, and the contours increase in intervals of 2$\sigma$.  The solid oval shows the fit used to derive the velocity gradient in the source from the SO$_2$ emission.} 
\label{fig:g10}
\end{figure}

\begin{figure}
\includegraphics[width=0.75\textwidth,angle=90]{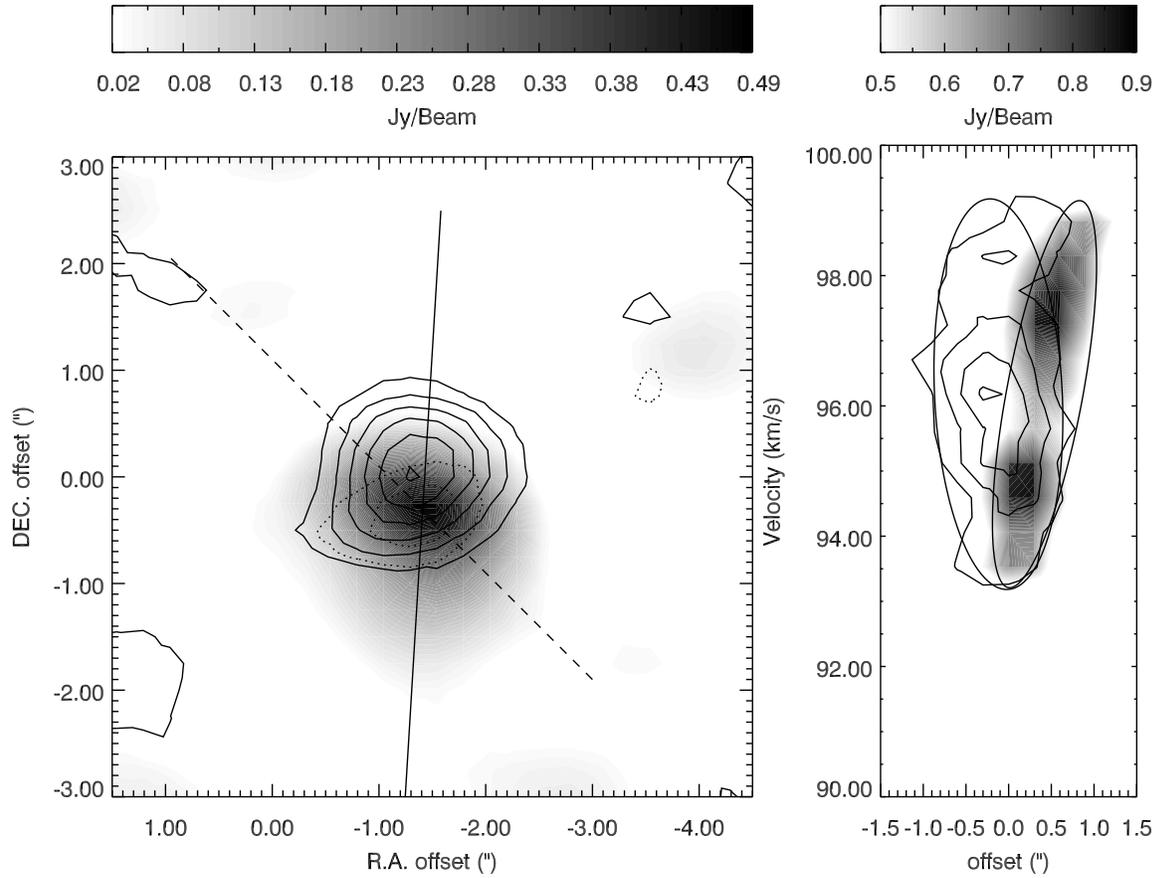}
\caption{{\bf (G28.2-0.04)}   This figure is similar to Figure \ref{fig:g10}, however the contours in the left panel start at 5$\sigma$ instead of 3.  The two ovals in the right panel show the fits to the SO$_2$ emission and the OCS emission. This is the only source for which the two species do not show the same velocity gradient.}
\label{fig:g28}
\end{figure}

\begin{figure}
\includegraphics[width=0.75\textwidth,angle=90]{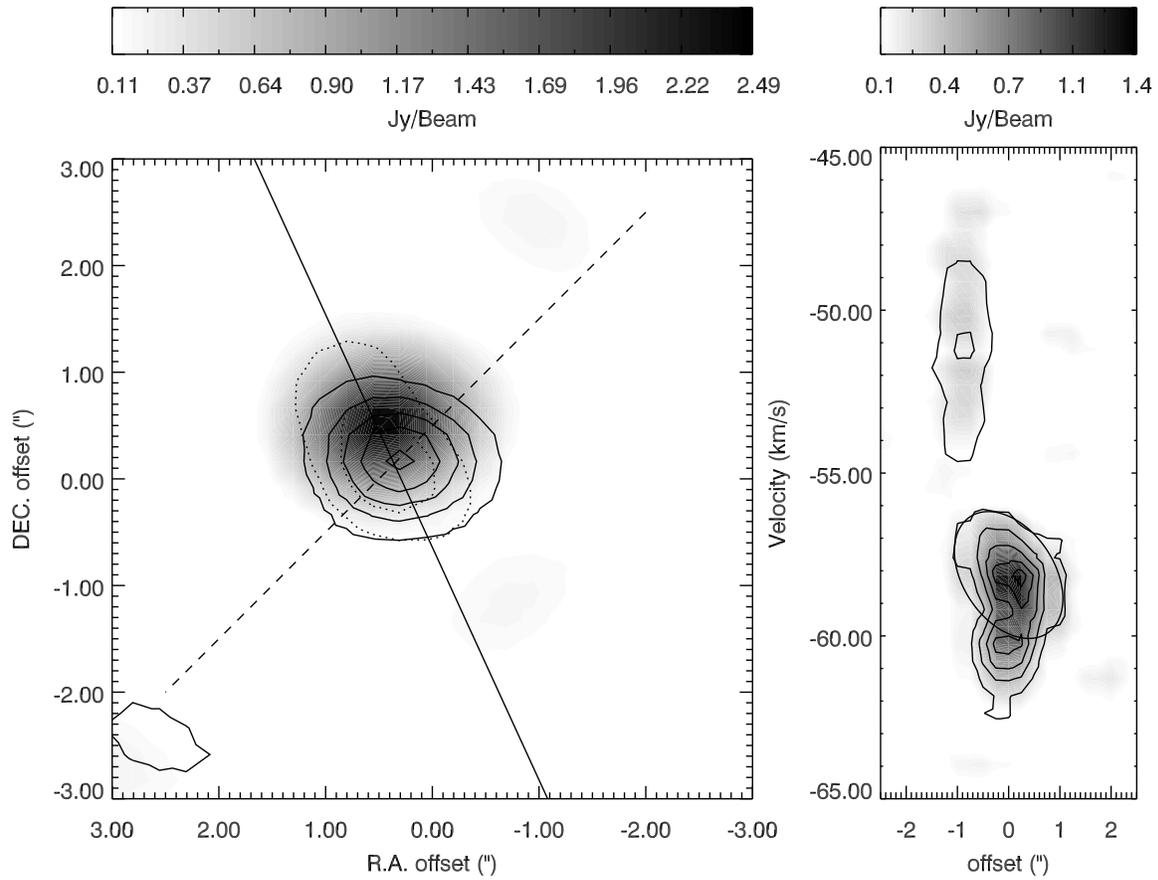}
\caption{{\bf (NGC 7538 IRS1)}   This figure is similar to Figure \ref{fig:g10}, however the contours in the left panel start at 10$\sigma$ instead of 3, and increase in intervals of 10$\sigma$ instead of 2.  In the right panel, the contours begin at 5$\sigma$ and increase in intervals of 5$\sigma$.}
\label{fig:n7}
\end{figure}

\begin{figure}
\includegraphics[width=0.75\textwidth,angle=90]{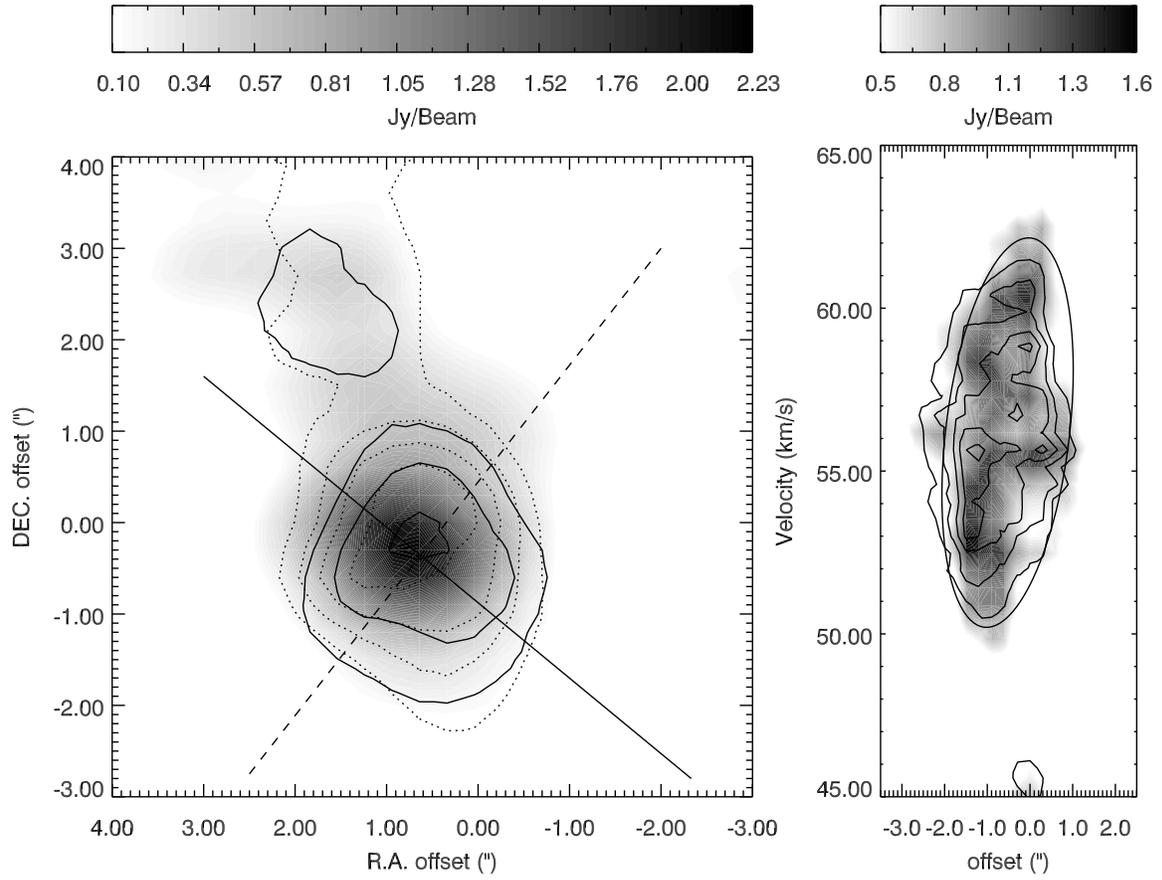}
\caption{{\bf (W51e2)}   This figure is similar to Figure \ref{fig:g10}, however the contours in the left panel start at 5$\sigma$ instead of 3, and increase in intervals of 4$\sigma$ instead of 2.}
\label{fig:w51e2}
\end{figure}

\begin{figure}
\includegraphics[width=0.75\textwidth,angle=90]{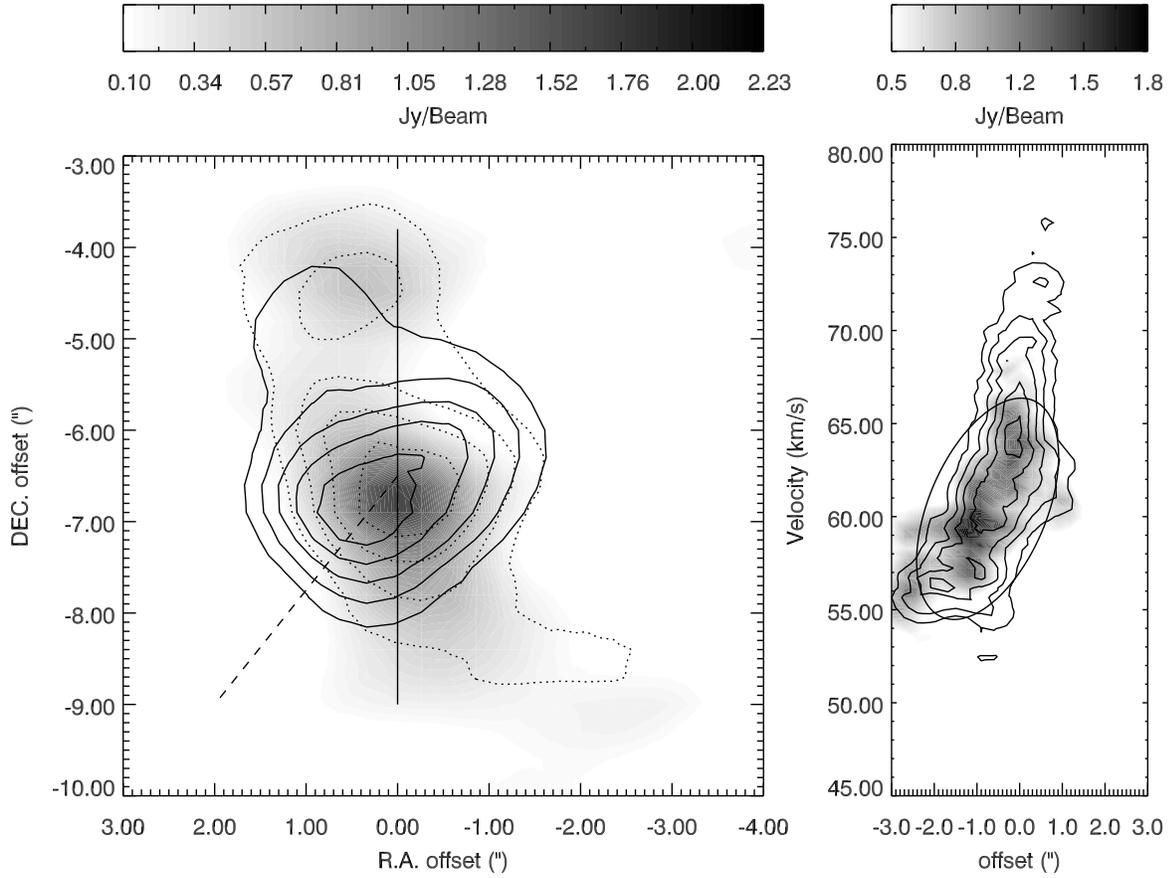}
\caption{{\bf (W51e8)}   This figure is similar to Figure \ref{fig:g10}, however the contours in the left panel start at 5$\sigma$ instead of 3,  increase in intervals of 4$\sigma$ instead of 2, and the oval represents the OCS emission instead of the SO$_2$ emission.  Note that only the red outflow lobe has been seen in CO for this source (Klaassen et al. in prep).}
\label{fig:w51e8}
\end{figure}

\begin{figure}
\subfigure[G10.6+0.4]{\includegraphics[width=0.4\textwidth,angle=-90]{fig6a.ps}}
\subfigure[G28.2-0.04]{\includegraphics[width=0.4\textwidth,angle=-90]{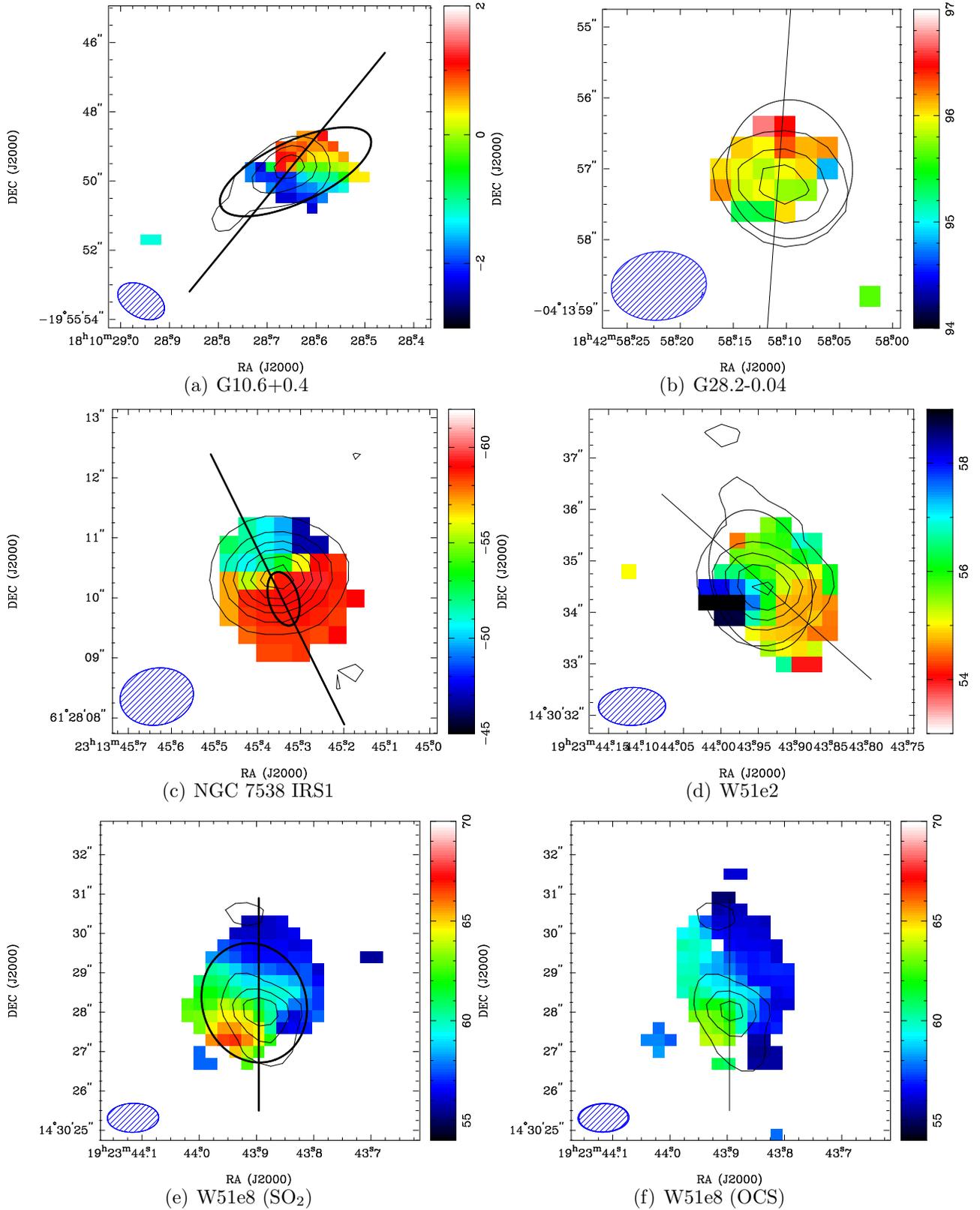}}
\subfigure[NGC 7538 IRS1]{\includegraphics[width=0.4\textwidth,angle=-90]{fig6c.ps}}
\subfigure[W51e2]{\includegraphics[width=0.4\textwidth,angle=-90]{fig6d.ps}}
\subfigure[W51e8 (SO$_2$)]{\includegraphics[width=0.4\textwidth,angle=-90]{fig6e.ps}}
\subfigure[W51e8 (OCS)]{\includegraphics[width=0.4\textwidth,angle=-90]{fig6f.ps}}
\caption{ First moment maps of SO$_2$. Lines show the direction of the velocity gradient, while thick ovals show the deconvolved SO$_2$ emitting regions (FWHM), and thin contours show the continuum emission starting at 5$\sigma$ and increasing in intervals of 5$\sigma$ (except for NGC 7538 where the contours start at, and are in intervals of 10$\sigma$).   Panels {\it e} and {\it f} show the SO$_2$ and OCS first moment maps for W51e8 (respectively).  Note that the OCS velocity gradient is in the same direction as that for SO$_2$, but does not show contamination from the outflow. }
\label{fig:rotation}
\end{figure}

 Towards each of our observed HII regions, SO$_2$ and OCS show evidence for a velocity gradient across the emission region (see Figures \ref{fig:g10}-\ref{fig:w51e8}). The solid black lines shown in the left hand panels of Figures \ref{fig:g10}-\ref{fig:w51e8} represent the cuts taken for the position velocity diagrams shown in the right hand panels.  The directions of the cuts were chosen along the position angles (PAs) with the largest velocity gradients, and perpendicular to the outflow direction if known.  The CO also observed as part of these datasets, when combined with observations of the larger scale structures (Klaassen et al. in prep), show outflows from each HII region.  The directions of these outflows are shown by dashed lines in Figures   \ref{fig:g10}-\ref{fig:w51e8}. The solid ovals presented in Figure \ref{fig:rotation} show the Full Width at Half Maximum (FWHM) of the SO$_2$ emission region deconvolved from the SMA synthesized beam.  In both SO$_2$ and OCS, the velocity gradients are present in similar size scales and velocity ranges.  In two of these sources (G10.6 and W51e2, shown in Figures \ref{fig:g10} and \ref{fig:w51e2} respectively), similar velocity gradients have previously been observed in NH$_3$ by \citet{Sollins05} and \citet{Zhang97} (respectively).  They suggest that rotation is responsible for the velocity gradients, and we adopt this interpretation here as well, and extend it to the rest of our sources.  Towards G10.6, the velocity gradient seen in NH$_3$ by \citet{Sollins05} has the same orientation and velocity limits as the observations presented here. The velocity gradient in NH$_3$ towards W51e2 seen by \citet{Zhang97} also has the same orientation as that presented here.    \citet{KK08} see a similarly oriented velocity gradient in the ionized gas (H53$\alpha$) within W51e2 as well.  We note that when the direction of the bipolar outflow emanating from these massive star forming regions is known (i.e. \citet{Davis98} for NGC 7538 IRS1 and \citet{KK08} for W51e2), the orientation of our observed velocity gradient is perpendicular to the outflow.   The right hand panels of Figures \ref{fig:g10}-\ref{fig:w51e8} show that the velocity gradients in SO$_2$ and OCS. The velocity ranges shown in Figures \ref{fig:g10}-\ref{fig:w51e8} are much larger than the velocity ranges shown in Figure \ref{fig:rotation}.  The velocity ranges used for the first moment maps in Figure \ref{fig:rotation} are smaller than those shown in the PV diagrams due to a clipping imposed on the data and to highlight the direction of the velocity gradient in each source.  The directions of the velocity gradients in SO$_2$ and OCS are consistent for a given source, save G28.2, and with the exception of W51e8 (see Section \ref{subsec:outflow_contam}), the velocity ranges for both species towards a single source are the same.  Below, we present our derivations for the velocity gradient in each source and our derived dynamical masses.  In Section \ref{subsec:outflow_contam} we present a caveat to the velocity gradient derivation for W51e8.
  
\subsection{Mass Estimates from Velocity Gradients}
\label{subsec:dynamical_mass}

To determine the velocity shifts and angular sizes required to calculate our velocity gradients, we use the methods presented in \citet{Zhang97}.  We extracted PV diagrams in both SO$_2$ and OCS along the solid lines shown in the left panels of Figures \ref{fig:g10}-\ref{fig:w51e8}. We then fit 2D Gaussians to the line emission. In each PV diagram, the Gaussians were fit to the 3 or 5 $\sigma$ level as stated in the figure captions.  The projected major and minor axes of these ellipses gave the angular extent and velocity shift across each source.  The calculated velocity gradients for each source are shown in Table \ref{tab:results} where the uncertainties arise from the velocity channel spacing (0.53 km s$^{-1}$) and spatial resolution (0.3$''$)
 of the observations.

It is interesting to note that the velocity gradients in G28.2 are differently oriented from each other in the PV diagram, yet result in the same dynamical mass.  While the change in velocity across the sources is the same for both species, the higher velocity gas is at different positions.  This is possibly due to the two species being slightly offset from each other as seen in the right hand panel of Figure \ref{fig:g28}.

If we assume that the material in these regions is in virial equilibrium, we can determine the dynamical mass of the enclosed region.  Because we cannot constrain the inclination angle of these rotating structures from our observations, we will assume that the velocity gradients are observed primarily `edge on' (i.e. the observed velocities are the true velocities of the gas) and thus our derived masses are lower limits to the true masses for structures which are inclined with respect to the plane of the sky. 

We find velocity gradients in the range of 92-175 km s$^{-1}$ pc$^{-1}$ and note that the values derived for a given source from SO$_2$ and OCS agree to within one or two times the quoted uncertainties, which is why only the SO$_2$ values are given in Table \ref{tab:results}. That these values are so similar suggest that SO$_2$ and OCS are tracing the same bulk motions.   The velocity gradients we present in Table \ref{tab:results} are comparable to, although generally slightly smaller than, those inferred for three other massive star forming regions (G24A1, IRAS 20126 and IRAS 18089) by applying the same analysis methods described here to the data presented by \citet{Beltran06}, \citet{Cesaroni05} and \citet{Beuther08} (respectively). These sources were observed at similar resolutions, but are, in general closer than the sources studied here, and do not show evidence for HII regions.  We find that our results are generally higher than the velocity gradient found in the low mass star forming region L1251 \citep{Lee07} again using the same analysis on their presented velocity gradients.  Our derived velocity gradients result in dynamical masses of between 12 and 364 M$_\odot$.  For all sources except for our unresolved source (NGC 7538) the derived dynamical masses are large enough to account for one (or more) massive star and surrounding HII region. Because the dynamical mass accounts for all of the mass within its outer emitting region (i.e. the hot gas and dust inside the HII region as well as the warm molecular gas responsible for the emission), we expect these dynamical masses to be the largest values presented here.

\subsection{Outflow Contamination in SO$_2$ towards W51e8}
\label{subsec:outflow_contam}

In Figure \ref{fig:w51e8} we show both an emission map (left panel) and a PV diagram (right panel) for SO$_2$ and OCS in W51e8. In the right panel, the grayscale shows the OCS emission,  and the contours show the SO$_2$ emission.   Here, we see that the velocity gradient in SO$_2$ around W51e8 extends to higher red shifted velocities than blue (from the rest velocity of the source given in Table \ref{tab:observe}).  The solid ellipse shows the OCS emission. We suggest that the SO$_2$ emission redward of 66 km s$^{-1}$ is not related to dynamical rotation, and we suggest that the SO$_2$ velocity gradient is being contaminated by outflow emission. This excess red shifted emission is not observed in OCS (see Panel f of Figure \ref{fig:rotation}); however the orientation of the velocity gradient in OCS is consistent with that of the SO$_2$. If we blindly derive a dynamical mass for this region from the full velocity gradient in SO$_2$, we estimate an enclosed mass of 930 M$_\odot$. Interestingly the redshifted lobe of the bipolar outflow (as seen in CO, Klaassen et al., in prep.) from this source is located near the redshifted part of the rotation signature. It is not clear why there is no corresponding CO blueshifted outflow lobe, which makes it hard to determine a concrete outflow direction.   The correlation in both space and velocity between the SO$_2$ emission and CO emission suggests we are seeing outflowing material pushing through the SO$_2$ shell of warm gas surrounding the region.  The detection of H30$\alpha$ in this region is marginal at best, and while it has been detected as a continuum source at 1.3 cm \citep{Zhang97}, it has not been detected at 3.6 cm (down to an rms limit of $\sim$ 1.2 mJy beam$^{-1}$ \citep{Gaume93}).  These results bring into question whether W51e8 is an HII region.  Given the large mass of this region however, if it is not an HII region, we suggest that it is a very young object which is  transitioning from hot core to HII region.

If this high velocity redshifted excess is due to the outflow pushing through the warm SO$_2$ emitting region, as suggested by our data, then the molecular outflow observed towards this source is continuing to  be powered from within the HII region itself.  If outflows are driven by accretion, this result suggests that accretion is continuing onto the star(s) present in this HII region. It is unlikely that this red extension in the emission could come from a remnant type outflow \citep[see, for instance][]{msc} since, by the time the HII region would have grown onto observable scales, the outflow mechanism from within would have stopped; there would be nothing pushing the SO$_2$ outwards. We suggest that the  rotation signatures of the other sources are not contaminated by outflow since the outflows from those sources are perpendicular to the velocity gradients present in SO$_2$ and can be fit well with a Keplerian orbit centered on the local standard of rest velocity of the sources with no red or blue excesses as seen towards W51e8.

\section{Gas, Dust and Stellar Masses}
\label{sec:other_mass}

In parallel to the dynamical mass estimates from the SO$_2$ and OCS lines, we can estimate the mass of molecular gas responsible for the emission (Section \ref{subsec:emission_mass}) from the integrated intensities of the lines.  This calculation requires knowing the temperature of the gas in the emitting region. The gas temperature for each region was independently derived from the K ladders of the J=12-11 transition of CH$_3$CN.  From the K ladders, we also estimate a gas mass as described below. 

At 230 GHz, there are two main contributors to the measured continuum emission: a thermal dust component and a component from the free-free emission from the HII region.  In Section \ref{subsec:continuum_mass} we discuss how the two components were separated and how a gas mass was derived from the thermal dust component.  In Section \ref{subsec:stellar_mass}  we discuss how a minimum stellar mass for each region was derived from the free-free emission from each source.

\subsection{Temperature estimates from CH$_3$CN emission}
\label{subsec:temperature}

\begin{figure}
\includegraphics[width=0.75\textwidth,angle=-90]{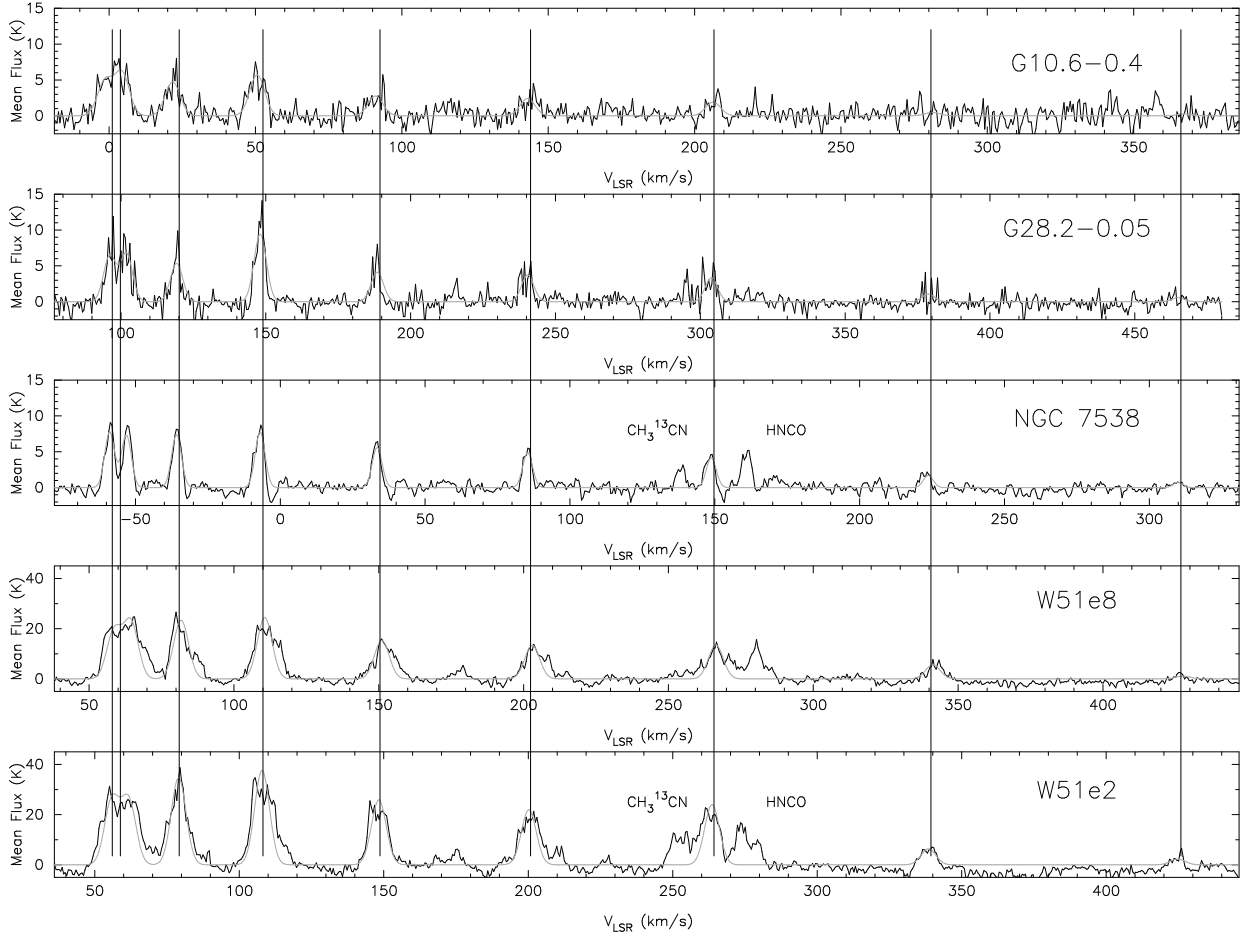}
\caption{CH$_3$CN integrated intensities and gaussian line fits for each source.  The vertical lines denote the expected velocities of each K component, and the velocities plotted along the x axis are with respect to the local standard of rest velocity. The plots have been aligned according to the rest velocity of the sources listed in Table \ref{tab:observe}.  The HNCO and CH$_3^{13}$CN lines near the K=6 component of the CH$_3$CN line have been marked in the middle and  bottom panels to show that they are contributing little contamination to the integrated intensity of the K=6 transition. Note that in NGC 7538, an inverse P-cygni profile can be seen in each K component suggesting that the CH$_3$CN may be tracing an infalling component of the gas.}
\label{fig:ch3cn_gauss_fits}
\end{figure}

CH$_3$CN is a symmetric top molecule, and as such can be used to trace the temperature in its emitting region.  The individual K levels within a specific J transition (in this case, J=12-11) are radiatively decoupled, meaning that the different K levels are populated through collisions alone \citep{Solomon71}, and the relative populations of the K components give an estimate of the ambient(kinetic) temperature through the rotational temperature diagram method described in \citet{Goldsmith99}.  We defined the emitting region for CH$_3$CN as twice the area of the FWHM of the integrated intensity of the combined emission from the K=0-8 components.  The CH$_3$CN spectra shown in Figure \ref{fig:ch3cn_gauss_fits} are averaged over this masked region. In order to determine the integrated intensity of each component in a systematic way, we followed the procedures  outlined in \citet{Remijan04}, \citet{Araya05} and \citet{Purcell06} which assume local thermodynamic equilibrium and that the gas is optically thin.  Gaussian profiles were fit to each component, keeping the widths and relative rest velocities fixed for a given source. This procedure allows the fitting routines to find the best fit to the rest velocity of CH$_3$CN in each source. As highlighted in the middle and bottom panels of Figure \ref{fig:ch3cn_gauss_fits}, HNCO and CH$_3^{13}$CN were within our spectral window but not fit by Gaussians.


From the fitted Gaussians to each K component, we determined the column density of the upper level which is related to the integrated intensity of the line through

\begin{equation}
N_u=\frac{8\pi k \nu^2}{A_{u\ell}hc^3}\int Tdv
\end{equation}

\noindent where $N_u$ is the column density in the upper level of the K component, $\nu$ is the frequency of the transition, $A_{u\ell}$ is the Einstein A coefficient for that transition, and $\int Tdv$ is the integrated intensity in units of K km s$^{-1}$.  $N_u$ is related to the total column density for the emitting region through:

\begin{equation}
\frac{N_u}{g_u}=\frac{N_{\rm TOT}}{Q(T)}e^{-E_u/kT}
\label{eqn:col_den}
\end{equation}

\noindent where $g_u$ is the degeneracy of the upper state, $Q(T)$ is the partition function, $N_{\rm TOT}$ is the total column density of the species, and $E_u$ is the energy above the ground state of the transition \citep[see the appendix of ][]{Araya05}.  Taking the log of both sides of the equation gives $ln(N_u/g_u)\propto -E_u/kT$. The ambient temperature can then be determined from the slope of the line of best fit.  

\begin{figure}
\includegraphics[width=0.75\textwidth,angle=-90]{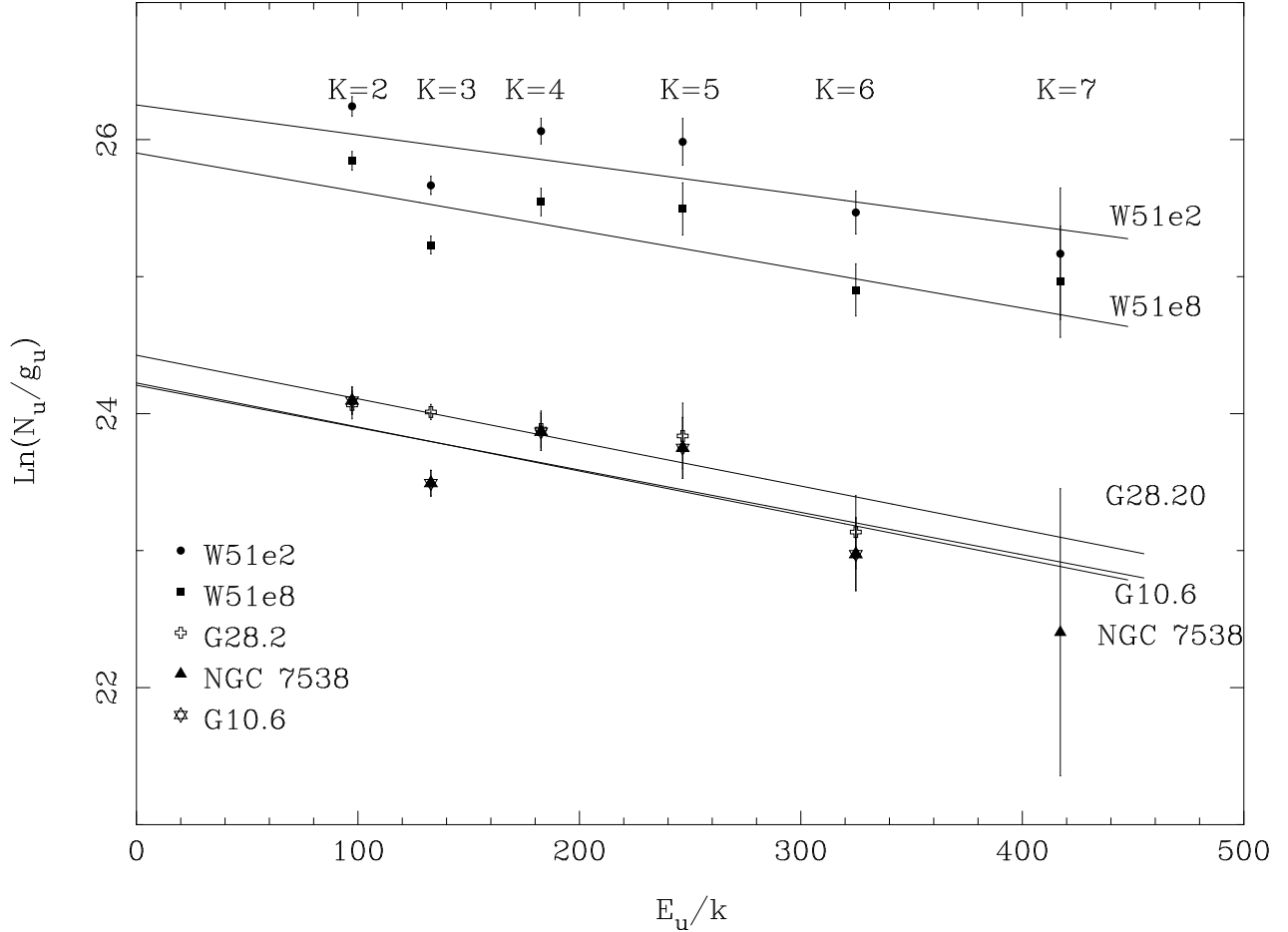}
\caption{Plot of the natural logarithm of the column density in the J=12-11 transition of CH$_3$CN divided by the degeneracy of the state as a function of the energy of the observed K component.  The inverse of the slope of the line of best fit shown here gives a measure of the temperature in each region.  The K=0,1 lines were blended in each spectrum, and thus the column densities for these components were not included in the fits.  For most sources, the integrated intensity of the K=8 component was below the noise level, and thus these values were also not included in the fits.}
\label{fig:temperature}
\end{figure}

In Figure \ref{fig:temperature} we show this plot for each of our regions and the derived temperatures are shown in Table \ref{tab:emission_mass}. Since these lines were observed simultaneously, the relative intensities of the lines are not affected by calibration uncertainties. The error bars for each data point come from the uncertainties in the Gaussian fits, and the uncertainties in the temperatures come from the uncertainty in the linear fit to the points in Figure \ref{fig:temperature}. The signal to noise ratios of the Gaussian fits to the K=8 components for  each source (as well as the K=7 components for G10.6 and G28.2) were not high enough to be included in the fits, or the plots presented in Figure \ref{fig:temperature}.  Also, due to the blending of the K=0,1 components they were also excluded from the linear fits presented in Figure \ref{fig:temperature}.  CH$_3$$^{13}$CN is detected in three of our sources.  The isotopic ratio of $^{12}$C and $^{13}$C is approximately 55 \citep{WR94}, so the detection of CH$_3$$^{13}$CN suggests our CH$_3$CN emission is optically thick.  Using Equation 1 of \citet{Choi93}, we find optical depths of $<$5.8, 50.4, 35.8, $<$8.9 and 16.3 for G10.6, G28.2, NGC 7538, W51e2 and W51e8 respectively.  The lower limits on the optical depths are for sources in which CH$_3$$^{13}$CN was not detected and the rms noise was used as an upper limit on the CH$_3$$^{13}$CN emission peak.  If the gas is optically thick, then we are overestimating the temperatures in our regions and underestimating the column density.

From these data, we find that the temperatures in the warm molecular gas are between 300 and 460 K.  These values are slightly higher than  the temperature range in which SO$_2$ is expected to be most abundant, which is consistent with our derivations giving upper limits to the temperatures.  \citet{Doty02} show that the abundance of SO$_2$ dramatically increases at 100 K, and then begins to decline again at temperatures greater than 300 K.  This lower temperature limit comes from the activation energy required for H$_2$ to combine with OH \citep{Doty02}.  At temperatures greater than $\sim$ 300 K this reaction becomes much less efficient \citep{Charnley97}.  The temperature range over which OCS has a high abundance is unclear; however, from the observations of \citet{Hatchell98} it appears to have fairly similar abundances towards a number of hot cores.  \citet{Charnley97} show that after 10$^5$ yr, the abundance of OCS should fall as HOCS$^+$ is removed from the system (the primary source of OCS).

The peak velocities of the CH$_3$CN lines are consistent with the V$_{\rm LSR}$ for each source (given in Table \ref{tab:observe}), and have similar emitting regions to the SO$_2$ and OCS gas.  Thus, we suggest  the CH$_3$CN  is tracing the same gas as the SO$_2$ and OCS, and will use the temperatures derived here to estimate the column densities of OCS and SO$_2$.

\subsection{Mass Estimates from Molecular line Emission}
\label{subsec:emission_mass}

We can find, from Equation \ref{eqn:col_den} not only an upper limit to the  rotational temperature of the CH$_3$CN emitting region, but the column density of CH$_3$CN gas responsible for the emission.  The slope of the line of best fit gives the temperature of the region while the y-intercept gives $\ln({\rm N/Q(T)})$. Using equation A33 from \citet{Araya05} we can solve for the column density of CH$_3$CN assuming the temperatures derived above for the partition function. Using the optical depth correction technique described in \citet{Goldsmith99}, we calculated the function C$_\tau$ for the optical depths (see their Equations 16 and 22) and correct our total column density values accordingly.  We use a CH$_3$CN abundance of $2\times10^{-7}$ \citep[an average of the values presented in][]{Remijan04}, and we  note that other authors find lower abundances in Orion \citep[i.e. $10^{-8}-10^{-9}$][]{WWP94}, and that these abundances would increase our CH$_3$CN derived gas masses closer to those presented below from our analysis of SO$_2$ and OCS; however the abundances presented in \citeauthor{Remijan04} were derived based partially on observations of one of the sources observed here (W51e2).   We present the masses derived from the CH$_3$CN emission both assuming optically thin and optically thick emission in Table \ref{tab:emission_mass} (the second last, and last columns respectively).

From the integrated intensity of a single transition of a molecule (such as SO$_2$ or OCS) we can also determine the average column densities of the gas responsible for that emission. The column density of either SO$_2$ or OCS can be derived by solving Equation \ref{eqn:col_den} for $N_{\rm TOT}$.  The column density in the observed species can be converted into a molecular gas column density by scaling by the abundance of the species (here SO$_2$ or OCS) with respect to H$_2$.  

For both SO$_2$ and OCS, there are large variations in the abundances given in the literature.  By averaging the abundances of these two species from a number of sources \citep{Watt85,Leen88,blake94,Charnley97,Hatchell98,vandertak03,Viti04} we find SO$_2$ and OCS abundances of $1\times10^{-7}$ and $2.5\times10^{-8}$ (respectively).  The ranges in abundances for these two species were 5$\times10^{-8}-1\times10^{-6}$ for SO$_2$ and $1\times10^{-9}-5\times10^{-7}$ for OCS. The mass of each emitting region was determined by multiplying the average molecular column density by the area and by the mass of molecular hydrogen.  We find that, to within our uncertainty estimates, the masses derived from the SO$_2$  are approximately twice those derived from the OCS emission.  For both species, over estimating the temperature acts to over estimate the mass.   We find that in general SO$_2$ is observed at a higher signal to noise ratio than OCS, however, given that SO$_2$ begins to deplete out of the gas phase at temperatures greater than 300 K \citep{Doty02}, we may be overestimating the mass from the SO$_2$ emission by overestimating the abundance with respect to H$_2$. This could be artificially inflating our SO$_2$ derived emission masses. These derived masses represent only the warm gas surrounding each HII region, and do not reflect the hot material within the HII region, or the cold gas exterior to the hot core. Thus, we expect these masses to be lower than the dynamical masses which include the stellar and ionized gas masses within the HII region.

\subsection{Mass Estimate from Continuum Emission}
\label{subsec:continuum_mass}

The two main components of the continuum emission in our sources are free-free emission from the HII region, and the warm dust from its surroundings. The predicted emission from the free-free component of the continuum flux can be calculated via  equation 2.124 of \citet{Gordon02}.  This free-free emission can then be removed from our observed 1.4 mm continuum, and we assume that the remaining continuum emission is due to thermal dust.

Using the equations and relations presented in \citet{hildebrand83}, we determined the dust masses in the regions producing the dust continuum emission.   We have extrapolated the dust emissivity from \citet{hildebrand83} as $Q_\lambda = 7\times 10^{-4}(0.125/\lambda_{mm})^\beta$, where 0.125 and $\lambda_{mm}$ are both wavelengths in units of mm (here, $\lambda_{mm}=1.4$).  This expression was derived without assuming that $
\beta = 1$ at 250 $\mu$m.   For our calculations, we assume a gas to dust ratio of 100 which allows for the conversion from dust mass to gas mass. We further assume that the gas and dust temperatures are not well coupled within the HII region, since the gas temperature within the HII region is well above the dust sublimation temperature.  Instead, we assume that the central star is heating the dust, and the temperature falls off as $T_{\rm dust}=(R_*/R)^{0.4}\times T_*$ \citep{Lamers_book}.  For the region in which the temperature is below the dust sublimation threshold, the temperature gradient was then averaged over our 1$''$ beam (or, a linear size of 6400 AU at the average distance to our sources) to give an average temperature of 400 K. The masses listed in Table \ref{tab:results} were calculated using this single temperature and assuming $\beta$ = 1.5. Increasing $\beta$ to 2 (while keeping the temperature at 400 K) increases the derived masses by more than a factor of 3. If we use the gas temperatures derived from our CH$_3$CN the changes in the masses derived from the dust emission are within one sigma of the values listed in Table \ref{tab:results}.

We find, in general, that the dust masses derived in this manner (and presented in Table \ref{tab:results}) are lower than the derived dynamical masses presented in Section \ref{subsec:dynamical_mass}.  In fact, we find that the average dust derived mass is approximately half of the average dynamically derived mass. This result is expected since our calculated dust masses are for regions which are not quite as extended as the SO$_2$ emitting region and do not reflect the masses of the central star(s) as is the case for the dynamical masses.

\subsection{Stellar Mass Estimates}
\label{subsec:stellar_mass}

Using the spectral energy distribution fits to the continuum data presented in \citet{KZK08} (both with and without density gradients within the HII region) we estimate the masses of the central stars powering the observed HII regions.  From their continuum measurements, the ionizing flux required to maintain the HII region was determined and a minimum stellar mass inferred.   If the HII region were expanding, the stellar mass required to produce the necessary ionizing flux would be higher. The required minimum stellar masses to maintain these HII regions (assuming density gradients within the HII region) are presented in Table \ref{tab:results} for all sources except W51e8 which was not included in the analysis of \citeauthor{KZK08}.

As stated above, we find that the average source mass derived from the continuum emission is approximately half of the dynamically derived mass.  When the stellar mass is added to the dust mass, we find that the combined mass is within approximately 1 $\sigma$ of the dynamical mass for G10.6 and G28.2.  For NGC 7538, the combination of dust and stellar mass is much greater than the dynamical mass; hence the estimate of 12 M$_\odot$ for the dynamical mass must be an underestimate.  For W51e2, we find that the dynamically derived mass is much greater than the combined dust and stellar mass estimates, which suggests that perhaps the dust derived mass maybe underestimated.

\section{Discussion and Conclusions}
\label{sec:discussion}

Using SO$_2$ and OCS emission, we have detected rotational signatures in the warm molecular gas surrounding five HII regions.  The bulk rotation of the gas inside the HII region \citep[as seen within W51e2 by][]{KK08} and around them (as shown here) suggests that each of the HII regions may have formed from a single collapsing core \citep[i.e.][]{RM08}.  If these rotational motions are indicative of a smaller scale disk within each HII region, these disks may be similar to the photoevaporating disks within hypercompact HII regions described in \citet{Keto07}.

Here we have presented two methods for determining the gas masses of the hot cores surrounding each HII region: from the molecular line emission and from the thermal component of the continuum emission.   Generally, the SO$_2$ emission mass is larger than the dynamical mass, however, as stated earlier, we may be over estimating the SO$_2$ abundance at the high temperatures we derive for our sources (which are upper limits due to optical depth effects), and our dynamical masses are lower limits since we do not know the inclination angles of the individual sources. We find that the average of the derived warm gas masses (SO$_2$ and OCS for a given source), when combined with the average minimum stellar mass required to power the HII region, are a significant fraction of the average derived dynamical masses.  Note that for NGC 7538, the stellar and gas masses are much larger than the dynamical mass, suggesting the dynamical mass is underestimated for this unresolved source.  That the gas and stellar masses are a large fraction of the dynamical mass in each resolved region suggests that these regions may be in rough dynamical equilibrium.  While we find that most of the continuum emission in each source is due to free-free emission, it is interesting to note that the expected dust contribution to the continuum emission is the highest in W51e8, a region we suggest to be quite young and only beginning to form its HII region.  In our other four sources, the dust emission contributes less than half of the continuum emission, while in this source it contributes at least 72\%  of the emission.  Other authors have also suggested that this region may be a mostly dusty source \citep{Lai01}.

NGC 7538 is the only source in this sample in which the deconvolved SO$_2$ emitting region and continuum source are not resolved by the SMA beam. It is also the only source where the dynamical and gas masses are too small to be responsible for forming the massive star(s) suggested by the stellar masses presented in Table \ref{tab:results}.  The outer radius of the maser disk seen by \citet{Pestalozzi04} in this region is $\sim$ 1000 AU, also smaller than our resolution element.  That the emission is unresolved greatly increases the uncertainty in our mass estimates. 

We note that previous studies have found much higher masses for the G10.6 system \citep[i.e. a few hundred solar masses,][]{KW06}, however the masses we derive here appear to be internally self consistent.  In contrast, the cores in W51e appear to be much more massive than G10.6. This may be due to the active hot cores surrounding the HII regions in these sources enhancing both the chemical abundances and radii of the sources; each of which act to increase our mass estimates.  

We find the masses derived for the two regions in W51e to be quite large.   The dynamical mass of W51e2 derived by \citet{Zhang97} is 10 M$_\odot$ on size scales of 0.3$''$.  Our value of 364 M$_\odot$ within 1.2$''$ is consistent with their mass projected out to the size scales we observed ($M_2 = M_1*(V_2^2r_2/V_1^2r_1)$), given the uncertainties in the measurements.  The large masses associated with these two regions suggests that they must be forming massive clusters of stars, as has suggested to be the case for G10.6 as well \citep[i.e.][]{Sollins05}.

As discussed earlier, there appears to be a component of the SO$_2$ emission in the W51e8 region which may be due to outflow.  That we detect outflowing gas in the warm layers surrounding the HII region suggests that the mechanism driving the outflow must be enclosed within the shell of SO$_2$ emission; it likely originates somewhere inside the HII region.  This supports the suggestion that there is ongoing accretion onto this (these) massive protostar(s).

\begin{deluxetable}{lcccccccr@{$\times$}lcr@{$\times$}lcc}
\rotate
\tablecolumns{14}
\tablewidth{0pt}
\tablecaption{Observational Properties}
\tablehead{
\colhead{Source} & \multicolumn{4}{c}{Coordinates\tablenotemark{a}} &
\colhead{V$_{\rm LSR}$\tablenotemark{b}} & \colhead{Dist.} & \colhead{rms\tablenotemark{c}} &
\multicolumn{3}{c}{Beam} &  \multicolumn{3}{c}{Source Size\tablenotemark{d}} \\
\cline{2-5	}\cline{9-11}\cline{12-14}
 &RA & DEC & $\Delta \alpha$& $\Delta \delta $  &&&& \multicolumn{2}{c}{maj.$\times$min.} & \colhead{PA}& \multicolumn{2}{c}{maj.$\times$min.} & \colhead{PA} \\
 &(hms)&($^\circ\, '\,''$)&$('')$&$('')$& \colhead{(km s$^{-1}$)} & \colhead{(kpc)} &\colhead{(Jy/Beam)} & \colhead{($''$)}&\colhead{($''$)} & \colhead{($^\circ$)}& \colhead{($''$)}&\colhead{($''$)} & \colhead{($^\circ$)} 
}
\startdata
G10.6-0.4 	& 18 10 28.7 &-19 55 49 		&11 &15			& -3 	& 6.0 & 0.32   	& 1.5&0.9&61.0	&4.8&1.8&-64.9	\\
G28.2-0.04	& 18 42 58.17 &-04 13 57.0	&-1.5&-0.3			& 97 	& 9.0 & 0.19   	& 1.3&1.0&-84.5	&1.5&1.1&-16.0	\\
NGC 7538 	& 19 13 22.069& 10 50 52.5	&0.3&0.3			& -59 	& 2.7 & 0.12  	& 1.2&0.9&-79.3	&0.9&$<$0.9&15.8	\\
W51e2 		& 19 23 43.913 &14 30 14.7 	&0.6&-0.3			&55 	&5.1  & 0.23   	& 1.3&0.7&-88.6	&2.8&1.9&17.5	\\
W51e8		& 19 23 43.913 &14 30 14.7 	&0&-6.6			&55	& 5.1 & 0.23  	& 1.3&0.7&-88.6	&3.1&2.6&19.5	\\
\enddata
\tablenotetext{a}{Coordinates given are for the pointing center of the observations; the offsets show the source center with respect to the pointing center.}
\tablenotetext{b}{V$_{\rm LSR}$ values taken from: \citet{Purcell06} (G10.6), \citet{Sollins05_g28} (G28), \citet{Zheng01} (NGC 7538 IRS1), \citet{Sollins04} (both W51e regions).} 
\tablenotetext{c}{RMS noise limits refer to channel spacings of 0.53 km s$^{-1}$ at 220 GHz.}
\tablenotetext{d}{In SO$_2$ emission deconvolved from synthesized beam.}

\label{tab:observe}
\end{deluxetable}

\begin{deluxetable}{lccccccc}
\rotate
\tablecolumns{8}
\tablewidth{0pt}
\tablecaption{Dynamical Masses, Dust Derived Masses and Minimum Stellar Masses}
\tablehead{
 &   \multicolumn{3}{c}{SO$_2$}  && \multicolumn{2}{c}{Dust derived Masses} &\colhead{Stellar Masses}\\
 \cline{3-5}
\colhead{Source} & \colhead {P.A\tablenotemark{a}}&
\colhead{Radius\tablenotemark{b}} &\colhead{Vel. Gradient} & \colhead{Mass} &
\colhead{S$^{\rm dust}_{\rm 1.4 mm}$} & \colhead{Mass}\\
 &\colhead{($^\circ$)}&\colhead{($''$)} & \colhead{(km s$^{-1}$ pc$^{-1}$)} & \colhead{(M$_\odot$)} &  \colhead{(Jy)} &\colhead{(M$_\odot$)} &\colhead{(M$_\odot$)}\\
}
\startdata
G10.6-0.4 	&140&1.5	&92$\pm$15 	  & 162$\pm$27 		&2.0$\pm$0.2 & 136$\pm$14 &69 \\
G28.2-0.04 	&173&0.45	&153$\pm$26 	  & 68$\pm$20\tablenotemark{c}			&0.6$\pm$0.2 & 95$\pm$34 & 34\\
NGC 7538 	&223&1	&156$\pm$44  & 12$\pm$3   		&2.5$\pm$0.7 & 36$\pm$10 & 62\\
W51e2 		&51&1.5	&175$\pm$22    & 364$\pm$51 		&3.9$\pm$1.5 & 140$\pm$53 & 87\\ 
W51e8		&0&1.6	&157$\pm$25	  & 320$\pm$45 		&1.8$\pm$0.7 & 82$\pm$33 & \nodata\\
\enddata
\tablenotetext{a}{Position angle of the cut for the PV diagram.}
\tablenotetext{b}{Estimated radius of the rotating region. Uncertainties are estimated to be half a pixel (0.15$''$).}
\tablenotetext{c}{Dynamical mass estimate for this source using OCS emission is 41$\pm$17 M$_\odot$.}
\tablecomments{Listed uncertainties are due to measurement error, not calibration uncertainty, which we estimate at 20\%}
\label{tab:results}
\end{deluxetable}

\begin{deluxetable}{lccccc|c|cccc}
\rotate
\tablecolumns{11}
\tablewidth{0pt}
\tablecaption{Temperatures and Molecular Emission Derived Gas Masses}
\tablehead{
\colhead{Source} &T$_{\rm kin}$\tablenotemark{a}&\multicolumn{4}{c}{SO$_2$} &\multicolumn{1}{c}{OCS} &\multicolumn{3}{c}{CH$_3$CN}\\
 & &Vel. Range\tablenotemark{c} & $\int T dv$ &N$_{SO_2}$ & M$_{\rm H_2}$\tablenotemark{b} &M$_{\rm H_2}$\tablenotemark{b} & $\ln(N/Q(T))$ & N$_{\rm CH_3CN}$ & M$_{\rm H_2}$\tablenotemark{b}&M$_{\rm H_2}^{\tau>1}$\\
 & (K)&(km s$^{-1}$) &(K km s$^{-1}$) & (10$^{15}$ cm$^{-2}$) & (M$_\odot$)  &(M$_\odot$)  & & (10$^{15}$ cm$^{-2}$) & (M$_\odot$)&(M$_\odot$)\\
 }
 \startdata
 G10.6-0.4	&323$\pm$105	&-6,4	&113$\pm$15&	37$\pm$4		&85$\pm$11	&57$\pm$3&24.2$\pm$0.2&1$\pm$0.3	&1$\pm$0.3	&5.6$\pm$1.6	\\
 G28.2-0.04	&300$\pm$89	&90,100	&175$\pm$27&	52$\pm$7		&131$\pm$20	&65$\pm$6&24.4$\pm$0.1&1$\pm$0.7	&1$\pm$0.6	&47$\pm$28	\\
 NGC 7538	&310$\pm$94	&-65,-45	&297$\pm$12&	92$\pm$3		&7.6$\pm$0.6	&3.1$\pm$0.4&24.2$\pm$0.2&1$\pm$0.6	&0.2$\pm$0.1 & 5.9$\pm$3.6		\\
 W51e2		&460$\pm$130	&45,75	&799$\pm$58&	429$\pm$27	&525$\pm$45	&294$\pm$32&26.3$\pm$0.1&21$\pm$14	&16$\pm$11	&265$\pm$188	\\
 W51e8		&350$\pm$83	&45,75	&634$\pm$58&	232$\pm$18	&385$\pm$37	&220$\pm$27&25.9$\pm$0.1&8$\pm$4		&7$\pm$4		&66$\pm$34\\
 \enddata
 \tablenotetext{a}{Upper limit on gas temperature due to optical depth effects in the CH$_3$CN emission}
\tablenotetext{b}{SO$_2$, OCS, and CH$_3$CN abundances used to calculate the total gas mass were 1$\times10^{-7}$, 2.5$\times10^{-8}$, and 2$\times10^{-7}$ (respectively)}
\tablenotetext{c}{Listed velocity ranges were used to create the zeroth and first moment maps}
\tablecomments{Listed uncertainties are due to measurement error, not calibration uncertainty, which we estimate at 20\%}
 \label{tab:emission_mass}
 \end{deluxetable}

\acknowledgements

We would like to acknowledge the support of the National Science and Engineering Research Council of Canada (NSERC). We would also like to thank our anonymous referee whos insights greatly improved this manuscript.

\end{document}